\documentclass[letterpaper]{article}

\usepackage[T1]{fontenc}

\usepackage{geometry}
\usepackage{setspace}

\usepackage[style = chem-acs]{biblatex}
\usepackage{graphicx}
\usepackage{float}
\newfloat{scheme}{Hp}{los}
\floatname{scheme}{Scheme}
\floatname{chart}{Chart}
\newfloat{graph}{Hp}{loh}
\usepackage[utf8]{inputenc}

\usepackage[version = 4]{mhchem} 
\usepackage{physics}
\usepackage{supertabular}
\usepackage{amssymb}
\usepackage{xcolor}
\usepackage{soul}
\usepackage[normalem]{ulem} 
\usepackage{comment}

\usepackage{authblk}
\author[1]{Jason S. R. McCoombs}
\author[1]{Jorge I. Hilari}
\author[2]{Jérôme Robert}
\author[3]{Balwant Singh Chauhan}
\author[3]{Ratnamala Chatterjee}
\author[4]{Filippo Troiani}
\author[1]{Athanassios K. Boudalis*}
\affil[1]{Institut de Chimie UMR7177 (Université de Strasbourg-CNRS), Universitè de Strasbourg, 4 rue Blaise Pascal, CS 90032
F-67081 Strasbourg Cedex, France}
\affil[2]{Institut de Physique et Chimie des Matériaux de Strasbourg UMR7504 (CNRS-Université de Strasbourg), 23 rue du Loess, 67037, STRASBOURG Cedex 2, France}
\affil[3]{Physics Department, I.I.T Delhi, New-Delhi, Hauz Khas 110016, India}
\affil[4]{CNR-Istituto Nanoscienze, via Giuseppe Campi 213/a, I-41125 Modena, Italy}

\title{ Second-harmonic signal in electric-field-modulated \\ EPR spectra of Fe$_3$ spin triangles}
\date{*Email: bountalis@unistra.fr}

\begin{document}

\maketitle

\begin{abstract}
  We present electric-field-modulated electron paramagnetic resonance (EFM-EPR) measurements on centrosymmetric single crystals of the molecular spin triangle $[\mathrm{Fe}_3\mathrm{O}(\mathrm{O}_2\mathrm{CPh})_6(\mathrm{py})_3]\mathrm{ClO}_4\cdot\mathrm{py}$ ($\bf{Fe_3}$). We provide the first observation of second harmonic EFM-EPR signal in polynuclear magnetic molecules. This signal is simulated and explained in terms of an electric-field induced modulation of the isotropic exchange in the molecule, and of their symmetry lowering resulting from a Jahn-Teller effect. Additionally, an unexpected first harmonic EFM-EPR signal is observed. Various plausible symmetry-breaking mechanisms are discussed in an attempt to explain this feature, whose observation is unexpected in a nominally centrosymmetric crystal.  
\end{abstract}

\section*{Keywords}

\begin{enumerate}
    \item Magnetoelectric coupling
    \item Spin triangles
    \item Electric-Field-Modulated EPR
    \item Quantum information
\end{enumerate}

\section{Introduction}
\label{sec:Introduction}
Interest in magnetoelectric (ME) molecular materials has significantly increased during the past decade, in part due to the proposal of Loss and coworkers that molecular spin triangles could encode a new type of spin-based qubit, the so-called ``spin-chirality qubit''~\cite{Trif08a, Trif10a}. This would be inherently controllable by electric fields through its ME coupling, while promising protection from magnetic noise \cite{Troiani12a}. Apart from these applicative implications, the ME properties of oligonuclear exchange-coupled systems are a new territory for condensed matter physics. This has, until recently, mostly focused on magnetic impurities in diamagnetic hosts studied with Electron Paramagnetic Resonance (EPR) under applied electric fields~\cite{Mims1976linear}, or on extended magnetic systems \cite{tokura_multiferroics_2014}, an area in which electric-field EPR studies have been less abundant \cite{kita_electric_1979,maisuradze_magnetoelectric_2012,laguta_magnetoelectric_2020}. Oligonuclear spin systems represent an intermediate case, where exchange coupling plays a role, but the spin Hamiltonian is defined in a reduced space and can be exactly diagonalized. 

In methodological terms, electric-field EPR has proved to offer both high sensitivities and rich experimental signatures that help in the elucidation of diverse phenomena. Three variations of this method have been reported: Continuous-Wave (CW) EPR under static electric-fields, pulsed EPR under pulsed electric-fields, and Electric-Field Modulated EPR (EFM-EPR). The latter technique replaces the modulated magnetic field utilized in traditional CW-EPR with a modulated electric-field, which changes the resonance conditions via the ME effect. The phase-sensitive detection of the technique allows for access to weak signals with experimentally accessible electric-fields, permitting the study of magnetically condensed samples that might not be amenable to pulsed techniques \cite{Perfetti26a}.

However, the applicability of this technique to centrosymmetric systems has so far been limited, because of the expected cancellation of the first-harmonic signal. In fact, it was suggested that first-harmonic EFM-EPR signals from nominally centrosymmetric systems could be attributed to the spurious magnetic fields created by the displacement currents between the electrodes~\cite{tacconi_sensitive_2026,song_exploring_2025}, rather than to actual ME effects. Here, we delve into this problem and provide two possible contributions. On the one hand, we show that a first-harmonic signal can indeed be obtained from a nominally centrosymmetric system. While no conclusive explanation for the presence of this signal is provided, we are able to exclude stray magnetic fields as a confounding factor. On the other hand, we measure a second-harmonic signal, for which we provide a detailed and quantitative physical interpretation in terms of a molecule deformation model.

\begin{figure}[H]
    \centering
	\includegraphics[width=0.4\columnwidth]{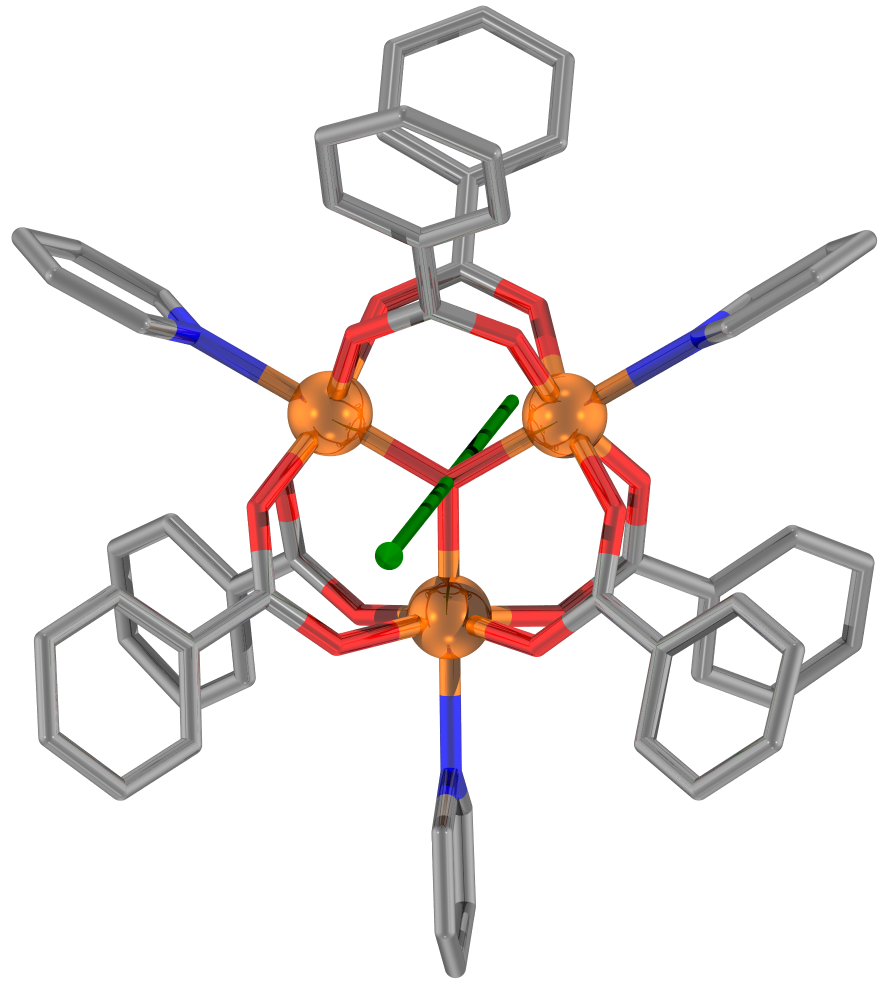}
	\caption{Pov-RAY plot of the cation of $\bf{Fe_3}$. The green arrow indicates the crystallographically-imposed $C_3$ axis.}
	\label{fig:Structure}
\end{figure}

The system under consideration consists of inversion-related spin triangles with formula
\newline $[\mathrm{Fe}_3\mathrm{O}(\mathrm{O}_2\mathrm{CPh})_6(\mathrm{py})_3]\mathrm{ClO}_4\cdot\mathrm{py}$ ($\bf{Fe_3}$, Figure \ref{fig:Structure}). A previous study of ours has highlighted apparent discrepancies between the crystallographic symmetry of the $\bf{Fe_3}$ molecule emerging from X-ray crystallography (space group ${\it{P}} 6_3/m$ down to 4.5 K) and the non-equilateral magnetic symmetry evidenced by EPR experiments~\cite{georgopoulou_dynamic_2017}. Moreover, our recent dielectric studies, based on direct phase-sensitive detection of the ME effect, have revealed clear first-harmonic ME signals with respect to the modulated magnetic field \cite{Singh2025Electrical}, although such signals should not be observed based on symmetry arguments alone.

\section{Materials and methods}
\label{sec:Materials_and_Methods}
\subsection{Sample preparation and experimental setup}
\label{subsec:Fe3_single_crystals}
$\bf{Fe_3}$ was synthesized as previously reported \cite{Boudalis18a}. Large single crystals were grown from slow cooling of a supersaturated solution in pyridine. The crystal habit was a regular, hexagonal dipyramid, with the crystallographic $c$-axis running along the dipyramid's $C_6$ axis. 

$X$-band EPR and EFM-EPR spectra were collected on an Elexsys E580 spectrometer fitted with a Bruker ER4118X-MD-5W dielectric resonator and a digital Signal Processing Unit. Multiharmonic detection was done up to the fifth harmonic. Low temperatures were achieved with an Oxford CF935 dynamic continuous flow cryostat, with the temperature regulated by an Oxford Mercury ITC servocontrol.

The sample holder was a parallel-plate capacitor, whose flat electrodes were painted on two parallel plastic surfaces with silver paste. The electrodes were connected with high-purity Ag wires to the core and sheath of a 50~$\Omega$ coaxial cable crimped with an SHV connector on its other end. Modulated voltage was supplied from the spectrometer's modulation cable and amplified through a home-made $\times25$ voltage transformer made from a ferrite core with 3 windings on the input and 75 windings on the output. The voltage amplitude and frequency were monitored in real time by an oscilloscope connected to the output of the transformer. The setup had been previously tested by measuring the voltage amplitude and phase at the electrodes and at the output of the modulation cable. Both voltages were found to be in phase. The voltage value was controlled by Xepr through the modulation amplitude setting. For a typical value of 400~V$_{pp}$, and with an inter-electrode distance of 2 mm, the electric-field strength was $2 \times 10^5~ \text{V/m}$ (peak-to-peak).

The $\bf{Fe_3}$ sample consisted of two crystals of maximum dimensions of $\sim2$~mm each, mounted with their respective $c$-axes perpendicular to each other, and aligned parallel and perpendicular to the electric-field direction and in both cases perpendicular to the rotation axis. This allowed for four different experiments to be carried out at only two different rotations of the sample holder, with orientation selection in each case accomplished by the magnetic field value, permitted by the fact that the $g_\parallel$$\left(\vb{B_0}\parallel c\right)$ and $g_\perp$$\left(\vb{B_0}\perp c\right)$ positions are separated by 0.2~T at the $X$-band.

All experiments were carried out at 5.0~K and at MW frequencies of approximately 9.74~GHz. Typical experimental conditions are: $B_{\mathrm{mod}} = 5 ~\mathrm{G_{pp}}$ and $f_{\mathrm{mod}} = \omega_m/2\pi = 100 ~\mathrm{kHz}$ for EPR experiments; $V_{\mathrm{mod}} = 380 ~\mathrm{V_{pp}}$, $f_{\mathrm{mod}}=\omega_m/2\pi = 20 ~\mathrm{kHz}$, $P_{\mathrm{MW}} = 150 ~\mathrm{mW}$ for EFM-EPR experiments. Additional experimental conditions per dataset are given in Table~S1. 
\begin{figure}[H]
    \centering
	\includegraphics[width=0.4\columnwidth]{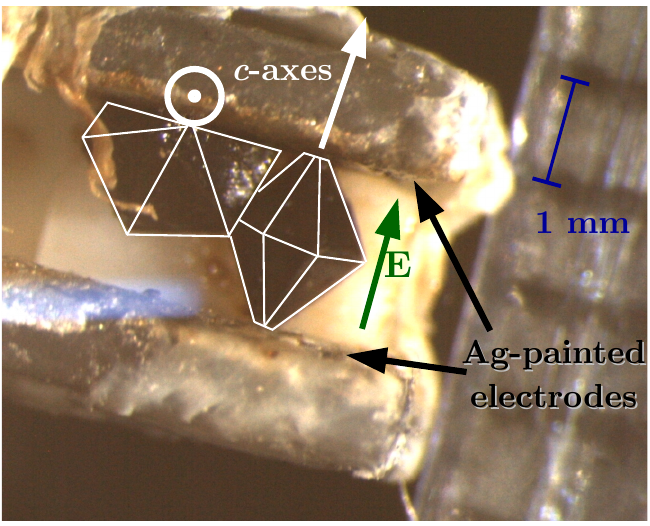}
	\caption{Photo of the crystals in the sample holder. The white outlines highlight the crystal faces and the white vectors indicate the crystallographic c-axes. The ruler graduations in the background are 1 mm apart.}
	\label{fig:Experimental_Setup}
\end{figure}

\subsection{Structural and magnetic symmetry analysis of Fe$_3$}
\label{subsec:Fe3}
\begin{figure}[H]
    \centering
	\includegraphics[width=0.8\columnwidth]{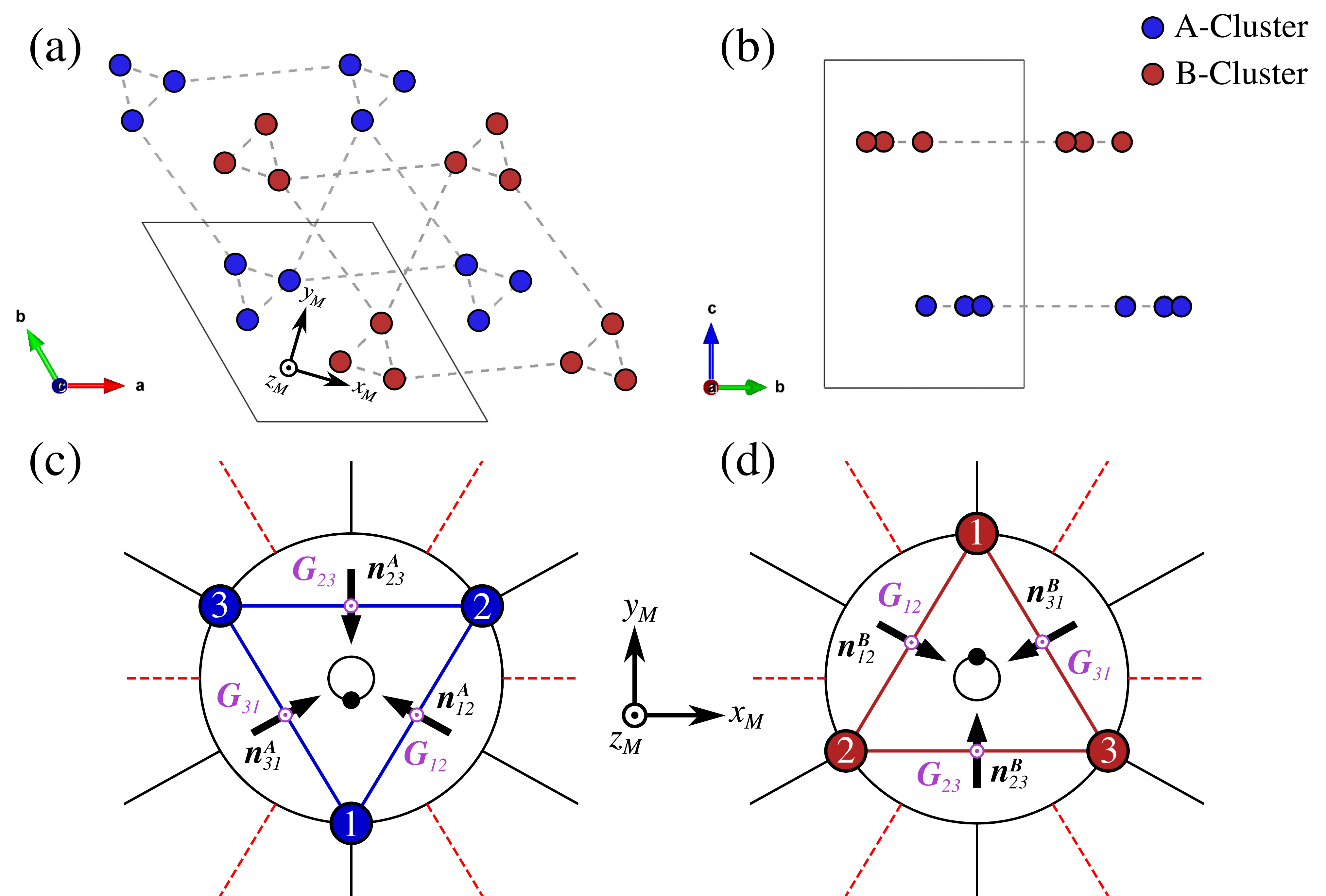}
	\caption{The simplified $\bf{Fe_3}$ crystal structure. (a) Shows the $\bf{Fe_3}$ triangles as viewed along the crystallographic $c$-direction and (b) the crystallographic $a$-direction. $\bf{Fe_3}$ $A$-clusters are shown in blue and inversion-related $\bf{Fe_3}$ $B$-clusters are shown in red. (c,d) A pair of inversion-related $A$ and $B$ clusters. The polarization unit vectors $\vu{n}_{ij}^\alpha$ and DM-vectors $\vb{G}_{ij}$ are indicated along with the pure isosceles- and scalene-type distortion angles depicted by the solid black and dashed red lines respectively. The $A$ cluster is shown with $\varphi=3\pi/2$ and the $B$ cluster is shown with $\varphi=\pi/2$, as indicated by the circles inside each triangle.
    }
	\label{fig:Fe3_Crystal_Simple}
\end{figure}

It is known from X-ray crystallography that the $\bf{Fe_3}$ system in the solid state preserves macroscopic inversion ($\mathcal{P}$) and three fold ($C_3$) symmetry down to 4.5~K. \cite{georgopoulou_dynamic_2017} In particular, one can define two types of crystallographic clusters, $A$ and $B$, which are inversion images of each other. These are arranged in different planes, as shown in a simplified representation of the single-crystal system in Fig.~\ref{fig:Fe3_Crystal_Simple}(a,b). 

However, it is also known that the $\bf{Fe_3}$ system breaks some of its macroscopic, crystallographic symmetries at the \emph{local} scale. The most obvious manifestation of this local symmetry breaking is the lowering of the $C_3$ symmetry in the spin Hamiltonian of a single triangular unit as deduced from CW-EPR measurements \cite{georgopoulou_dynamic_2017}. It is currently unknown if this distortion is static or dynamic in nature, but quantitative agreement with spectroscopic measurements was found by implementing a model in which 24 different, approximately equally populated, static triangular distortions coexist as an ensemble in the solid state \cite{georgopoulou_dynamic_2017}. The microscopic origin of this local symmetry reduction in various antiferromagnetic triangular systems has been hypothesized to be a Jahn-Teller-like distortion \cite{murao_jahn-teller_1974}. However, especially in $\bf{Fe_3}$, disordered freezing of pyridine solvates, ascertained by dielectric studies \cite{Singh2025Electrical}, could take place. This would lead to relatively strong, disordered, and local electric fields, $\vb{E}_{\mathrm{int}}(\vb{r})$, due to the large electric dipole moment residing in the pyridine molecules and primarily oriented in-plane (see Supporting Information). This disordered local field has the potential to alter the spin Hamiltonian of individual molecules via ME-coupling, as we will show, leading to a reduction in the $C_3$ symmetry of the unperturbed molecules. In theory, this effect could be working in concert with the Jahn-Teller distortion, as previously described~\cite{Azimi_Mousolou2016-lt}. 

\subsection{Theory of multi-harmonic EFM-EPR detection}
\label{subsec:Theory_of_EFM-EPR}
According to the above-mentioned dynamic model \cite{georgopoulou_dynamic_2017}, the isotropic exchange couplings between the spins $\vb{S}_i^\alpha$ and $\vb{S}_j^\alpha$ in the spin Hamiltonians of the $A$ and $B$ triangles read:
\begin{equation}
    J_{ij}^{\alpha,\zeta}(\varphi) = J\left\{1 + \eta_\zeta\left[\cos(\varphi -\vartheta_i^\alpha)+\cos(\varphi-\vartheta_j^\alpha)\right]\right\},
    \label{eq:J_of_varphi}
\end{equation}
where $\alpha = A,B$ defines the inversion partner, the angles $\left(\vartheta_1^A, \vartheta_2^A,\vartheta_3^A\right) = \left(3\pi/2, \pi/6, 5\pi/6\right)$ and $\vartheta_i^B = \vartheta_i^A +\pi$ define the positions of the magnetic ions in the spin triangles [Fig. \ref{fig:Fe3_Crystal_Simple}(c,d)], and $\varphi$ along with $\eta_\zeta$ specify the direction and magnitude of the symmetry lowering that affects the triangle. It should be noted that CW-EPR \cite{georgopoulou_dynamic_2017} as well as inelastic incoherent neutron scattering (IINS) \cite{Sowrey2001-oz} are both consistent with a model that includes a bimodal distribution of $\eta_\zeta$ values, with CW-EPR predicting $a_1 = 0.102$ and $a_2 = 0.118$. The physical origin of a bimodal distribution in $\eta_\zeta$ is still unclear. 

In addition to the $\varphi$-dependent isotropic exchange, it is also known that a sizable Dzyaloshinskii--Moriya (DM) antisymmetric exchange interaction (DMI) is present in this system~\cite{Rakitin1981-bv}. The fact that the triangular clusters are placed in crystallographic mirror planes restricts the DM vectors to point either parallel or anti-parallel to the $c$-axis of the material.
Likewise, the inversion symmetry
ensures that the DM vectors on both triangles should be identical $\left(\vb{G}_{ij}^A = \vb{G}_{ij}^B = \vb{G}_{ij}\right)$~\cite{Moriya1960-xc}. For a positive-valued DMI parameterized by $G_z$, the situation is shown in Fig.~\ref{fig:Fe3_Crystal_Simple}(c,d). Further, we neglect the single-ion anisotropy (zero-field splitting) and $g$-tensor anisotropy for  individual Fe ions. Previous calculations show in fact that, while finite, these terms would not contribute significantly to ME effects, aside from an overall shift of the energy levels \cite{leMardele25a}.  
The resulting zero-field spin Hamiltonian describing a single molecule of the ensemble is given by:
\begin{equation}
    \mathcal{H}^0_{\alpha,\zeta}(\varphi) = \sum_{k=1}^3J_{k,k+1}^{\alpha,\zeta}(\varphi)\,\vb{S}_k^\alpha\cdot\vb{S}_{k+1}^\alpha + G_z\vu{z}\cdot\sum_{k=1}^3\,\vb{S}_k^\alpha\times\vb{S}_{k+1}^\alpha\,.
\end{equation}

Magnetic-field-dependent terahertz absorption spectroscopy has confirmed that the $\bf{Fe_3}$ spin triangles also possess an in-triangular-plane spin-dependent electric polarization, which couples to the external electric field and allows a modulation of the hopping strengths in the super-exchange pathways in the spin triangle~\cite{Bulaevskii08a}. The operative electric-dipole operator capturing this behavior is defined as 
\begin{equation}
    \vb{p}_\alpha = \kappa\sum_{k=1}^3 \vu{n}_{k,k+1}^\alpha\left(\vb{S}_k^\alpha\cdot\vb{S}_{k+1}^\alpha\right),
    \label{eq:Polarization_operator}
\end{equation} 
where the $\vu{n}_{ij}^\alpha$ unit vectors are detailed in Fig.~\ref{fig:Fe3_Crystal_Simple}(c,d). The value of the magnetoelectric coupling was estimated to be $\kappa \approx 4\times10^{-4}$~$e\cdot$nm in Ref.~\cite{leMardele25a}, which is consistent with first-principles calculations. This estimate is also compatible with that of $\eta_\zeta$ derived from the simulation of CW-EPR spectra, if we combine Eq.~\ref{eq:J_of_varphi} with 
\begin{equation}
    J_{ij}^{\alpha,\zeta}(\varphi) = J + \kappa\vb{E}_\mathrm{int}\cdot \vu{n}_{ij}^\alpha,
\end{equation}
where $\vb{E}_\mathrm{int}$ corresponds to the random, internal electric fields resulting from the disordered pyridine sublattice, characterized by a bimodal distribution of magnitudes and preferential orientations of $n\pi/6$ (see the Supporting Information). The spin Hamiltonian of each cluster as a function of both $\vb{E}_\mathrm{ext}$ and $\vb{B}_0$ thus reads:
\begin{equation}
    \mathcal{H}_{\alpha,\zeta}(\varphi,\vb{E}_\mathrm{ext},\vb{B}_0) = \mathcal{H}^0_{\alpha,\zeta}(\varphi) -\vb{B}_0\cdot\vb*{\mu}_\alpha - \vb{E}_\mathrm{ext}\cdot\vb{p}_\alpha,
\end{equation}
where $\vb*{\mu}_\alpha = -g\mu_B\sum_{k=1}^3\vb{S}_k^\alpha$ is the magnetic moment of cluster $\alpha$ and $\vb{E}_\mathrm{ext}$ and $\vb{B}_0$ are externally applied electric and magnetic fields.

Importantly, the zero-field expectation value of $\vb{p}_\alpha$ is non-zero so long as $\eta_\zeta$ in Eq.~\ref{eq:J_of_varphi} is non-zero, and its orientation depends on the angle $\varphi$\footnote{Only in the case of an equilateral triangle does $\left<\vb{p}_\alpha\right>$ vanish, and even then, its fluctuations, $\left<\vb{p}_\alpha^2\right> - \left<\vb{p}_\alpha\right>^2$, remain finite due to non-zero electric-dipole mediated matrix elements between eigenstates. Thus, even in the equilateral triangle, the electric susceptibility tied to the spin-configuration remains important.}. Therefore, a finite in-plane electric polarization can be locally present on each molecule, not only because the individual triangular clusters lack inversion symmetry, but also because deformation lowers the $C_3$ symmetry. Further, the crystallographic inversion symmetry results in the following relations between expectation values of the $A$ and $B$ molecules at zero electric field: $E_n^{A\zeta}(\varphi,\vb{B}) = E_n^{B\zeta}(\varphi +\pi,\vb{B})$, $\vb*{\mu}_n^{A\zeta}(\varphi,\vb{B}) = \vb*{\mu}_n^{B\zeta}(\varphi + \pi,\vb{B})$, and $\vb{p}_n^{A\zeta}(\varphi,\vb{B}) = -\vb{p}_n^{B\zeta}(\varphi + \pi,\vb{B})$, where $\mathcal{O}_n = \bra{\psi_n}\mathcal{O}\ket{\psi_n}$ is the expectation value of the operator $\mathcal{O}$ in state $\ket{\psi_n}$. Inversion symmetry is recovered at the macroscopic level under the reasonable assumption that there are equal numbers of each crystallographic type ($A$ and $B$) and that the values of $\varphi$ are isotropically distributed (see below). A detailed investigation of the thermal expectation values of $\vb*{\mu}$, $\vb{p}$ and $E$ is included in the Supporting Information as a function of the distortion angle $\varphi$.

Applied electric fields can in principle affect different terms in the spin Hamiltonian of molecular nanomagnets. These terms include both the (isotropic and anisotropic) exchange interactions~\cite{Katsura2005-qs}, the $g$ tensors, and the single-ion anisotropies at each magnetic site~\cite{Ludwig1961-sr,Ham1961-te}. Hereafter, for the sake of simplicity and for consistency with previous investigations of ours, we focus on the modulation of the isotropic exchange interaction, resulting from the renormalization of the charge-transfer gaps in the ligands that mediate the super-exchange pathways~\cite{Bulaevskii08a,Trif08a}.

Since the excitation field in EFM-EPR experiments is a linearly polarized MW-frequency magnetic field, $\vb{B}_1(t) = \vb*{\mathcal{B}}_{1}\cos(\omega_1t)$, the relevant absorption spectrum depends on matrix elements involving the magnetic dipole operator ${\vb*\mu}$. The eigenstates of each individual cluster depend on all the Hamiltonian parameters, $\ket{\psi_n^{\alpha\zeta}(\varphi,\vb{B}_0,\vb{E}_\mathrm{ext})}$, and so do the eigenenergies and the Boltzmann factors $p^{\alpha\zeta}_{n}(\varphi,\vb{B}_0,\vb{E}_\mathrm{ext}) = e^{-E_{n}^{\alpha\zeta}(\varphi,\vb{B}_0,\vb{E}_\mathrm{ext})/k_BT}/\mathcal{Z}_{\alpha\zeta}$, with $\mathcal{Z}_{\alpha\zeta}=\sum_n e^{-E_{n}^{\alpha\zeta}(\varphi,\vb{B}_0,\vb{E}_\mathrm{ext})/k_BT}$. Thus, the microwave absorption induced by $\vb{B}_1$ in the presence of both a static magnetic field $\vb{B}_0$ and an electric field $\vb{E}_\mathrm{ext}$ is given by:
\begin{equation}
    \mathcal{A}(\omega_1;\vb{B}_0,\vb*{\mathcal{B}}_1,\vb{E}_\mathrm{ext}) \propto \sum_{\alpha=A,B} \sum_{\zeta=1,2} \sum_{i=1}^{12} q_{\alpha\zeta} (\varphi_i)\sum_{m<n} p^{\alpha\zeta}_{m}\left|\mel{\psi_n^{\alpha\zeta}}{\vb*{\mathcal{B}}_1\cdot\vb*{\mu}_\alpha}{\psi_m^{\alpha\zeta}}\right|^2 f(\omega_1,\omega^{\alpha\zeta}_{mn},\Gamma),
    \label{eq:General_Absorption_Fermi_Golden}
\end{equation}
where $q_{\alpha\zeta}(\varphi_i)$ is the probability density, relative to a molecule of type $\alpha$, for the deformation angle $\varphi=\varphi_i=i\pi/6$. The function $f(\omega_1, \omega^{\alpha\zeta}_{mn}, \Gamma)$ defines the line shape, which is modeled using Gaussians in this work, with $\omega_{mn}^{\alpha\zeta} = (E^{\alpha\zeta}_n-E^{\alpha\zeta}_m)/\hbar$, and characteristic width $\Gamma$. The electric field can affect the line shape by shifting the transition frequencies $\omega_{mn}^{\alpha\zeta}$, the Boltzmann factors, and the transition amplitudes. The line width $\Gamma$ is assumed to be independent of the molecule type and of the applied fields. Given the transition energies of the Fe$_3$ molecule and the temperature at which the measurements are performed, the absorption spectra are dominated by the transition between the Zeeman-split states belonging to the ground doublet.

In the case of a modulating electric field, $\vb{E}_\mathrm{ext} = \vb*{\mathcal{E}}\cos(\omega_m t)$, the general expression for the absorption will also be time-dependent and oscillatory with period $\mathcal{T}$. Therefore, the time-dependence of $\mathcal{A}$ can be conveniently captured by expanding in a Fourier series:
\begin{equation}
    \mathcal{A}(t;\omega_1,\vb{B}_0,\vb*{\mathcal{B}}_1,\vb*{\mathcal{E}}) = \sum_{n=0}^{\infty}A_n\cos(n\omega_m t) + B_n\sin(n\omega_m t),
    \label{eq:Fourier_Series}
\end{equation}
where the coefficients,
\begin{equation}
    \begin{aligned}
        A_n(\omega_1,\vb{B}_0,\vb*{\mathcal{B}}_1,\vb*{\mathcal{E}}) &= \frac{2}{\mathcal{T}}\int_{0}^\mathcal{T}\mathcal{A}(t)\cos(n\omega_m t) dt, \\
        B_n(\omega_1,\vb{B}_0,\vb*{\mathcal{B}}_1,\vb*{\mathcal{E}}) &= \frac{2}{\mathcal{T}}\int_{0}^\mathcal{T}\mathcal{A}(t)\sin(n\omega_m t) dt,
    \end{aligned}
    \label{eq:Harmonics}
\end{equation}
are interpreted as the $n$th-harmonic components of the in-phase and out-of-phase responses respectively~\cite{Boas2005-ky}. Therefore, in order to calculate the $n$th harmonic response, one first calculates the absorption over one full period of electric field modulation and then extracts the harmonic coefficients using Eq.~\ref{eq:Harmonics} through numerical integration.

\section{Results and discussion}
\label{sec:Results}
\begin{figure}[H]
    \centering
	\includegraphics[width=0.6\columnwidth]{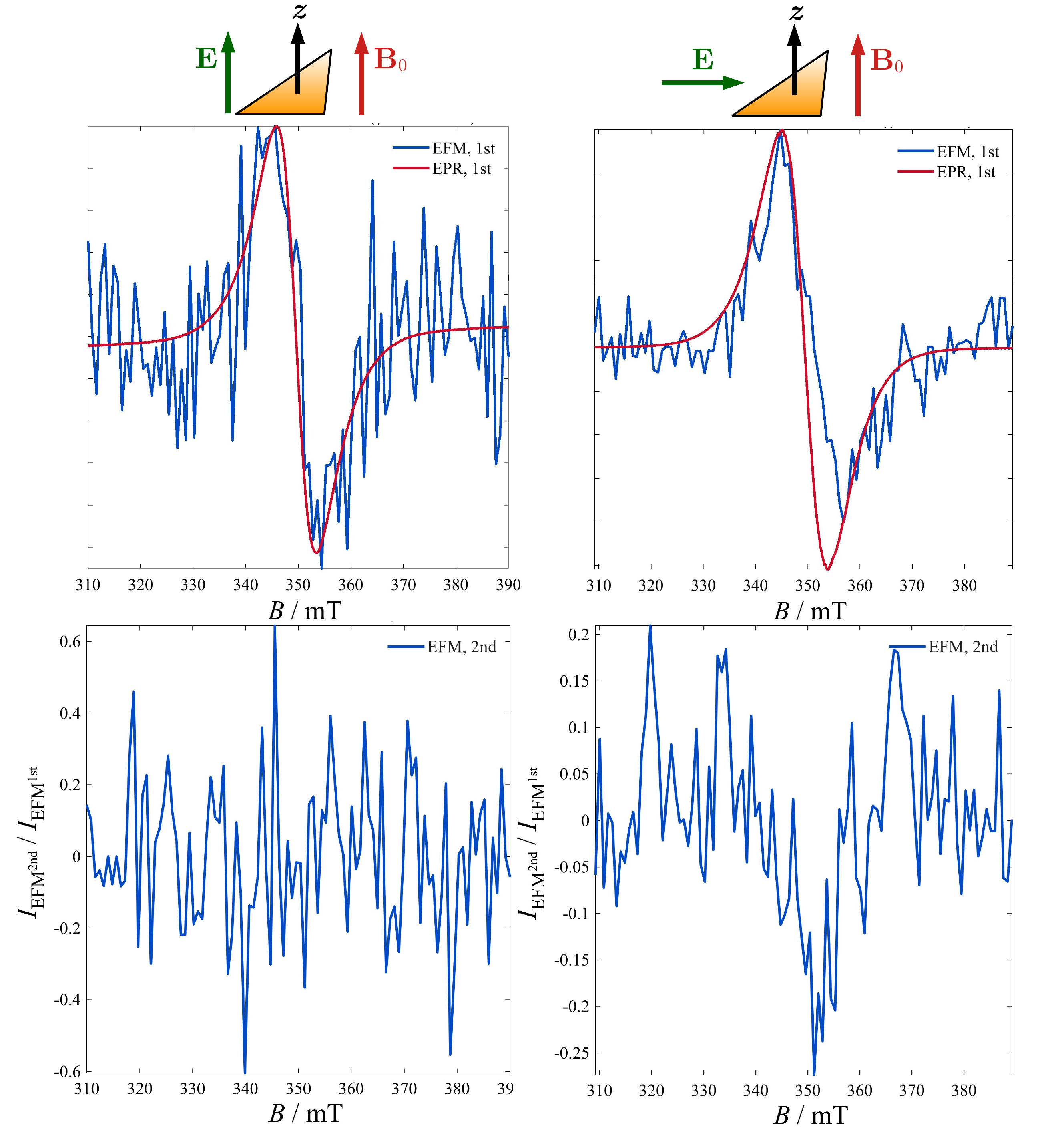}
	\caption{First (top) and second (bottom) harmonics of the EPR (red) and EFM-EPR (blue) signals of $\bf{Fe_3}$ at the $g_\parallel$ region. Extended experimental details are given in Table~S1}.
	\label{fig:g_parallel}
\end{figure}

\begin{figure}[H]
    \centering
	\includegraphics[width=0.6\columnwidth]{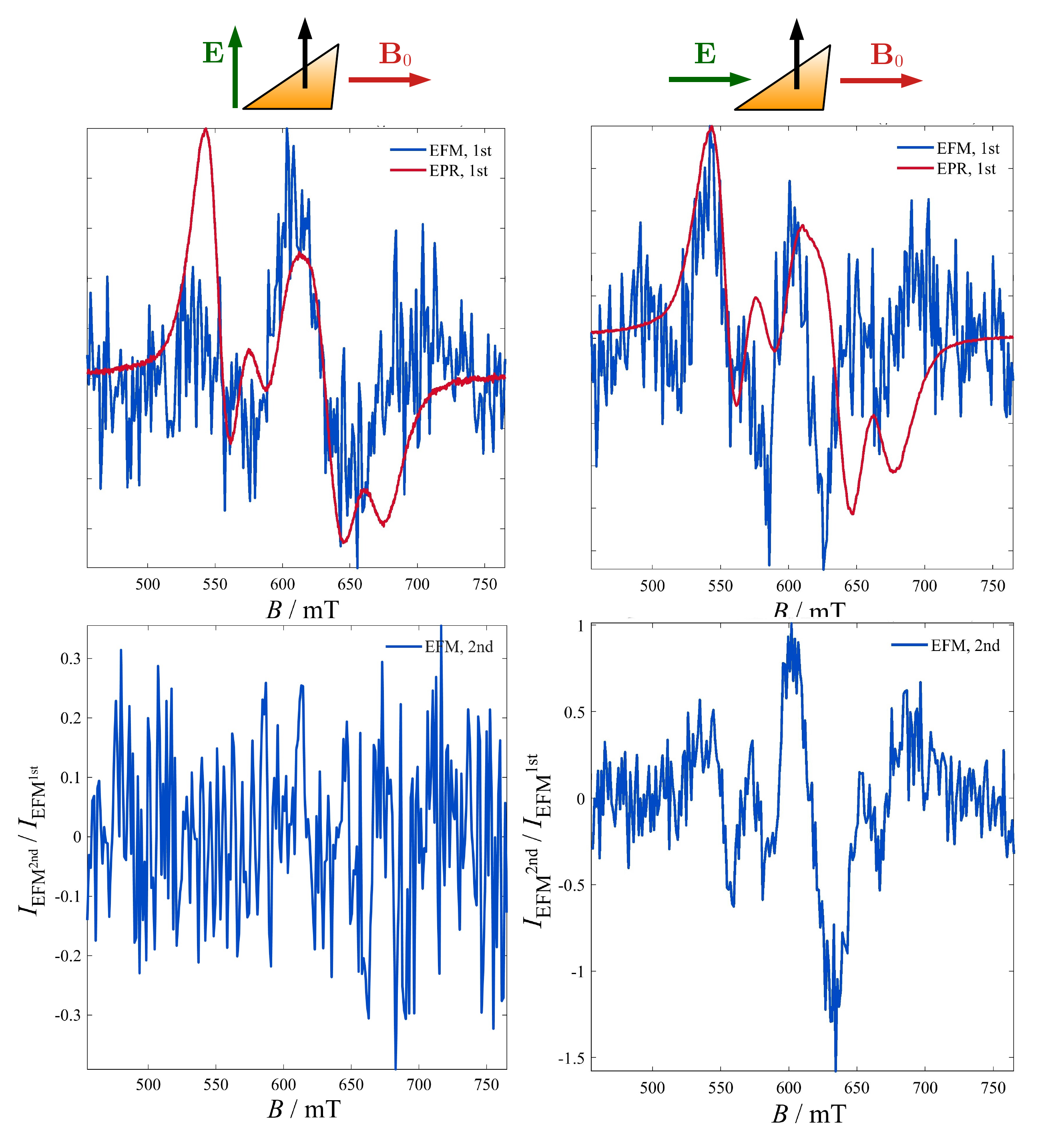}
	\caption{First (top) and second (bottom) harmonics of the EPR (red) and EFM-EPR (blue) signals of $\bf{Fe_3}$ at the $g_\perp$ region. Extended experimental details are given in Table~S1.}
	\label{fig:g_perpendicular}
\end{figure}

Standard EPR experiments reproduced the previously described single-crystal spectra  \cite{Boudalis18a} with remarkable fidelity, including the substructure of the $g_\perp$ ($\vb{B}_0\perp\vu{z}_M$) signal.
EFM-EPR experiments revealed clear first-harmonic signals for all four relative orientations of the samples with respect to $\vb{B}_0$ and $\vb*{\mathcal{E}}$. Interestingly, second-harmonic signals were recorded above the noise level only for the in-plane orientation of $\vb*{\mathcal{E}}$, \textit{i.e.} $\vb*{\mathcal{E}} \perp \vu{z}_M$. In those cases, the intensity of second-harmonic EFM signal was ca. 25 \%  of its first-harmonic counterpart for $g_\parallel$ ($\vb{B}_0\parallel\vu{z}_M$) and 100 \% for $g_\perp$. Experimental details on the series of measurements presented in Figs.~\ref{fig:g_parallel} and~\ref{fig:g_perpendicular} are listed in Table~S1.

\subsection{EFM-EPR simulations}
\label{subsec:Simulating}
\subsubsection{Longitudinal magnetic field and transverse electric field} 
\label{subsec:gparr}
\begin{figure}[H]
    \centering
	\includegraphics[width=1.0\columnwidth]{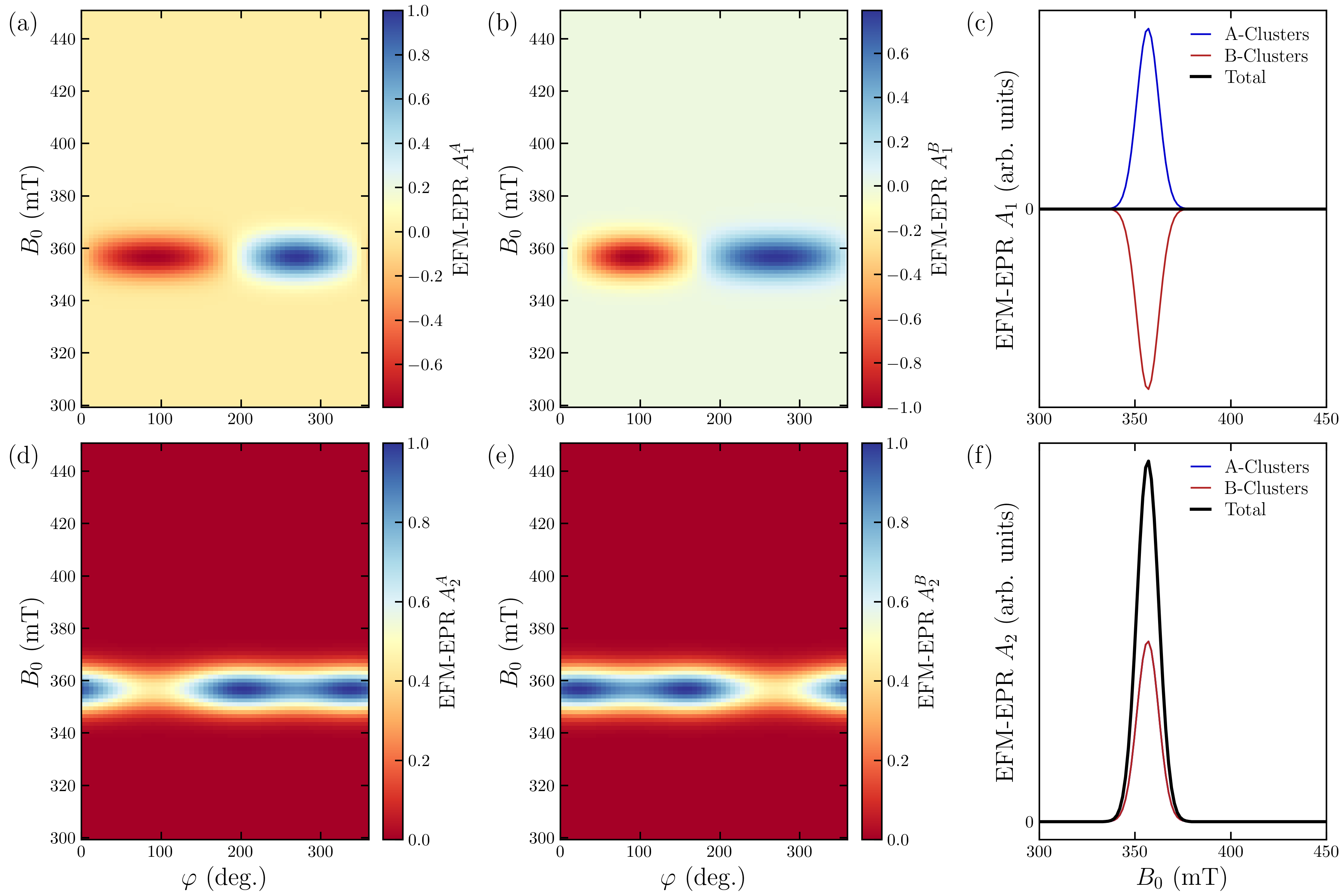}
	\caption{Numerical simulation of the first and second-harmonic signals, $A_1$ and $A_2$, of individual molecules with $\vb{B}_0\parallel\vu{z}_M$ and $\vb*{\mathcal{E}}\parallel\vu{y}_M$. (a) $A_1^A(B_0,\varphi)$ for $A$ clusters. (b) $A_1^B(B_0,\varphi)$ for $B$ clusters. (c) The ensemble averages $\left<A_1^A\right>_\varphi$, $\left<A_1^B\right>_\varphi$ and $\left<A_1\right>_{\alpha,\varphi}$. (d) $A_2^A(B_0,\varphi)$ for $A$ clusters. (e) $A_2^B(B_0,\varphi)$ for $B$ clusters. (f) The ensemble averages $\left<A_2^A\right>_\varphi$, $\left<A_2^B\right>_\varphi$ and $\left<A_2\right>_{\alpha,\varphi}$. In all cases, $\eta_\zeta = 0.110$, $\omega_1/2\pi = 9.74$~GHz, and $\Gamma/2\pi = 0.16$~GHz.}
	\label{fig:gparr_Ey_Bz_EFM}
\end{figure}

Before discussing the simulation of the experimental results and in order to provide a general understanding of the model, we report the first- and second-harmonic contributions from the molecules $A$ and $B$ as a function of $\varphi$. 
Figure \ref{fig:gparr_Ey_Bz_EFM} shows the EFM-EPR responses for $\vb{B}_0\parallel\vu{z}$ and $\vb*{\mathcal{E}}_1\perp\vu{z}$. For this orientation of the magnetic field, the resonance frequency, $\omega_{mn}^{\alpha\zeta} = \omega^{\alpha\zeta}_{01}$, responsible for the main observed line, is not affected by the oscillating electric field [Fig.~\ref{fig:Dispersion_vs_E}]. The first harmonic response (higher panels) results solely from the modulation of the transition amplitude and the Boltzmann factors. Even though the signal varies and changes sign as a function of $\varphi$ [Fig.~\ref{fig:gparr_Ey_Bz_EFM}(a,b)], nonzero values are obtained for each cluster averaging over $\varphi_i$. However, the averages take opposite values for the $A$ and $B$ molecules (assuming that $q_A(\varphi_i)=q_B(\varphi_j)=q$), so that the sum over the two vanishes [Fig.~\ref{fig:gparr_Ey_Bz_EFM}(c)]. The situation is different for the second harmonic contribution (lower panels), which varies but does not change sign as a function of $\varphi$ [Fig.~\ref{fig:gparr_Ey_Bz_EFM}(d,e)]. As a result, the average contribution is the same for the $A$ and $B$ molecules and the overall second-harmonic signal is nonzero [Fig.~\ref{fig:gparr_Ey_Bz_EFM}(f)]. 

\subsubsection{Transverse magnetic field and transverse electric field}
\label{subsec:gper}

\begin{figure}[H]
    \centering
	\includegraphics[width=1.0\columnwidth]{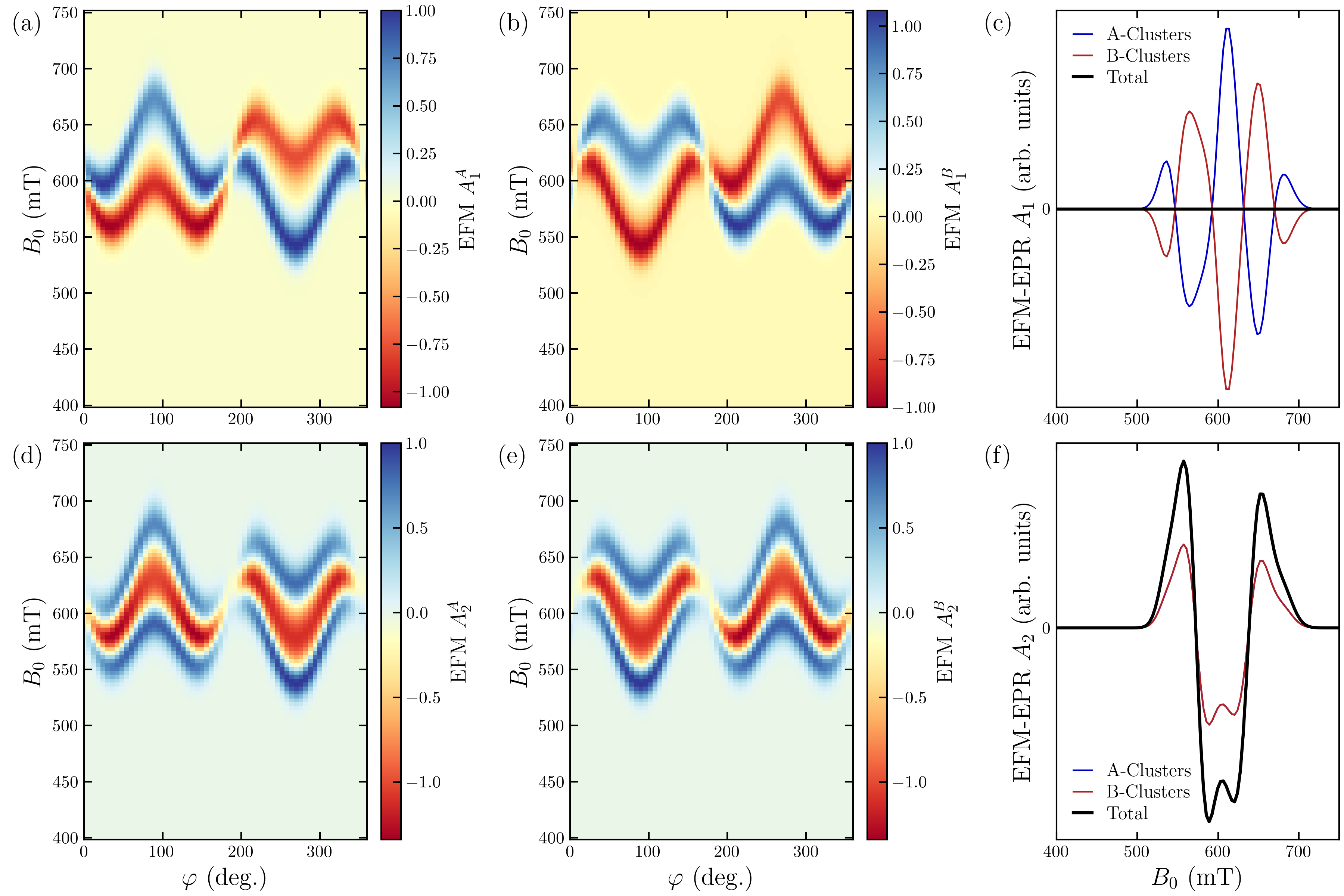}
	\caption{Numerical simulation of the first and second-harmonic signals, $A_1$ and $A_2$, of individual molecules with $\vb{B}_0\parallel\vb*{\mathcal{E}}\parallel\vu{y}_M$. (a) $A_1^A(B_0,\varphi)$ for $A$ clusters. (b) $A_1^B(B_0,\varphi)$ for $B$ clusters. (c) The ensemble averages $\left<A_1^A\right>_\varphi$, $\left<A_1^B\right>_\varphi$ and $\left<A_1\right>_{\alpha,\varphi}$. (d) $A_2^A(B_0,\varphi)$ for $A$ clusters. (e) $A_2^B(B_0,\varphi)$ for $B$ clusters. (f) The ensemble averages $\left<A_2^A\right>_\varphi$, $\left<A_2^B\right>_\varphi$ and $\left<A_2\right>_{\alpha,\varphi}$. In all cases, $\eta_\zeta = 0.110$, $\omega_1/2\pi = 9.74$~GHz, and $\Gamma/2\pi = 0.16$~GHz.
    }
	\label{fig:gparr_Ey_By_EFM}
\end{figure}

\begin{figure}[H]
    \centering
	\includegraphics[width=1.0\columnwidth]{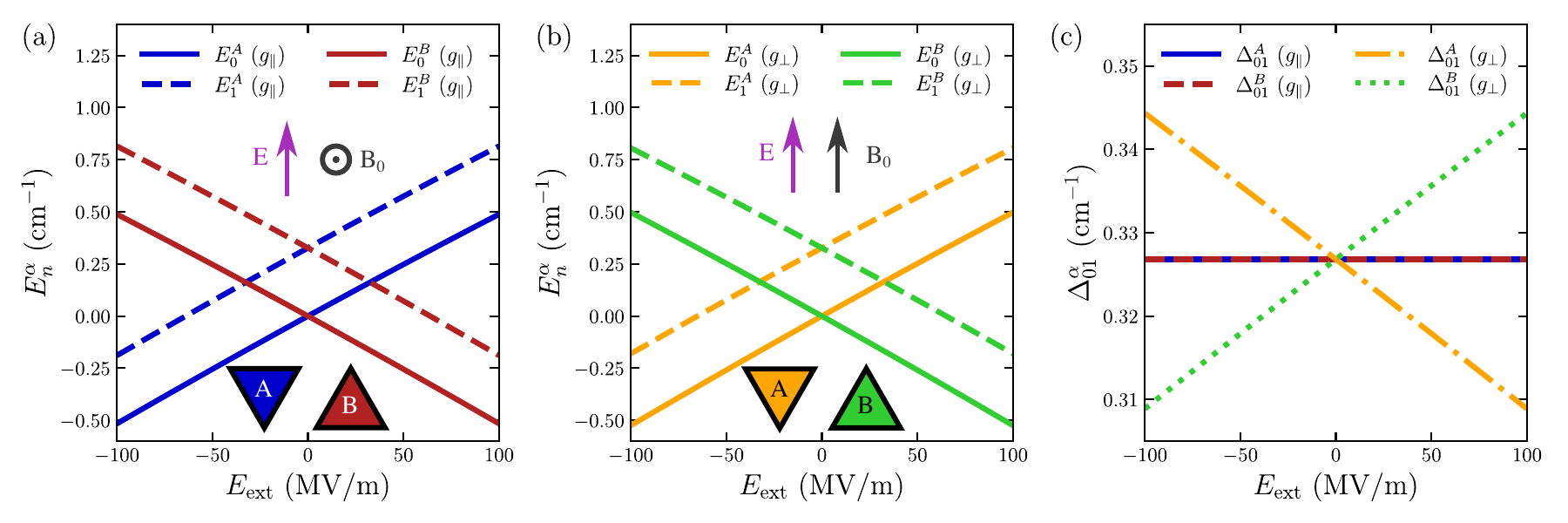}
	\caption{The effect of an applied electric field on the lowest lying Kramers doublet in $\bf{Fe_3}$. $E_{0,1}^A(\varphi=3\pi/2)$ and $E_{0,1}^B(\varphi=\pi/2)$ in the case of (a) $\vb{B}_0\parallel \vu{z}_M$ ($g_\parallel$) and (b) $\vb{B}_0\parallel\vu{y}_M$ ($g_\perp$). (c) The energy differences of the lowest energy Kramers pair $\left(\Delta_{mn}^{\alpha\zeta} = \hbar\omega_{mn}^{\alpha\zeta}\right)$, $\Delta_{01}^{A} = E_{1}^A(\varphi=3\pi/2) - E_{0}^A(\varphi=3\pi/2)$ and $\Delta_{01}^{B} = E_{1}^B(\varphi=\pi/2) - E_{0}^B(\varphi=\pi/2)$ for both the $g_\parallel$ and $g_\perp$ geometries. In the case of $g_\parallel$, $\vb{B}_0 = 350$~mT~$\parallel\vu{z}_M$, while for $g_\perp$, $\vb{B}_0 = 592$~mT~$\parallel\vu{y}_M$. For these calculations, $\eta_\zeta$ was chosen to be 0.118. In both cases, $\vb{E}_\mathrm{ext}\parallel\vu{y}_M$.}
	\label{fig:Dispersion_vs_E}
\end{figure}

A qualitatively different behavior is obtained for in-plane orientations of the magnetic field, and specifically for $\vb{B}_0\parallel\vu{y}$ and $\vb*{\mathcal{E}}_1\parallel\vu{y}$ (Fig.~\ref{fig:gparr_Ey_By_EFM}). Here, the oscillating electric field induces a modulation of the transition energies [Fig.~\ref{fig:Dispersion_vs_E}], whose sign and amplitude depend on the angle $\varphi$. This results in oscillations of the first-harmonic contribution as a function of the applied magnetic field, both for given values of $\varphi$ [Fig.~\ref{fig:gparr_Ey_By_EFM}(a,b)] and after averaging over such an angle [Fig.~\ref{fig:gparr_Ey_By_EFM}(c)]. However, as in the previous case, the contributions of an $A$ molecule at $\varphi$ are opposite to that of a $B$ molecule at $\varphi+\pi$, so that, for $q_A(\varphi_i)=q_B(\varphi_j)=q$, the contributions of the $A$ and $B$ molecules cancel each other [Fig.~\ref{fig:gparr_Ey_By_EFM}(c)]. As far as the second-harmonic is concerned, the contribution of an $A$ molecule at $\varphi$ coincides with that of a $B$ molecule at $\varphi+\pi$ [Fig.~\ref{fig:gparr_Ey_By_EFM}(d,e)], and this gives rise to a nonzero signal after averaging over the angle values [Fig.~\ref{fig:gparr_Ey_By_EFM}(f)]. 

One important observation is that the resonant field is far more sensitive to the applied electric field when the magnetic field is aligned along the triangular plane. In particular, while the simulated line-shapes for $\vb{B}_0\parallel\vu{z}$ are ``absorption-like'', those for $\vb{B}_0\parallel\vu{y}$ are ``derivative-like''. This is because, in the model, the electric field can modulate energy difference between the doublets, the magnetic-dipole transition amplitudes, and the $x$ or $y$ components of the effective $g$ factor, but not its $z$ component (and thus the energy of the dominant transition, for $\vb{B}_0\parallel\vu{z}$).

\subsubsection{Geometry-independent features}

The above results illustrate some symmetry-related features of the model, which apply to arbitrary orientations of the applied fields. In particular, the contribution to the first harmonic response from an $A$ cluster with deformation angle $\varphi$ is exactly opposite to that from a $B$ cluster with deformation angle $\varphi + \pi$. Conversely, the second harmonic response of an $A$ cluster with $\varphi$ is exactly equal to that of a $B$ cluster with $\varphi + \pi$. From this it follows that, if the probability distributions $q_\alpha(\varphi)$ of the deformation angles restore inversion symmetry at a statistical level (as is the case for $q_A(\varphi)=q_B(\varphi+\pi)$ and, more specifically, for constant distributions $q_A(\varphi)=q_B(\varphi)=q$), then the model predicts a vanishing first-harmonic signal and a nonzero second-harmonic response. The presence of a first-harmonic signal can only be explained by some mechanism that further lowers the symmetry of the molecules or that gives rise to a statistical unbalance between the $A$ and $B$ clusters (see below). 

\subsection{Quantitative comparison to experimental second harmonic spectra}
\label{subsec:Quantitative}
\begin{figure}[H]
    \centering
	\includegraphics[width=0.8\columnwidth]{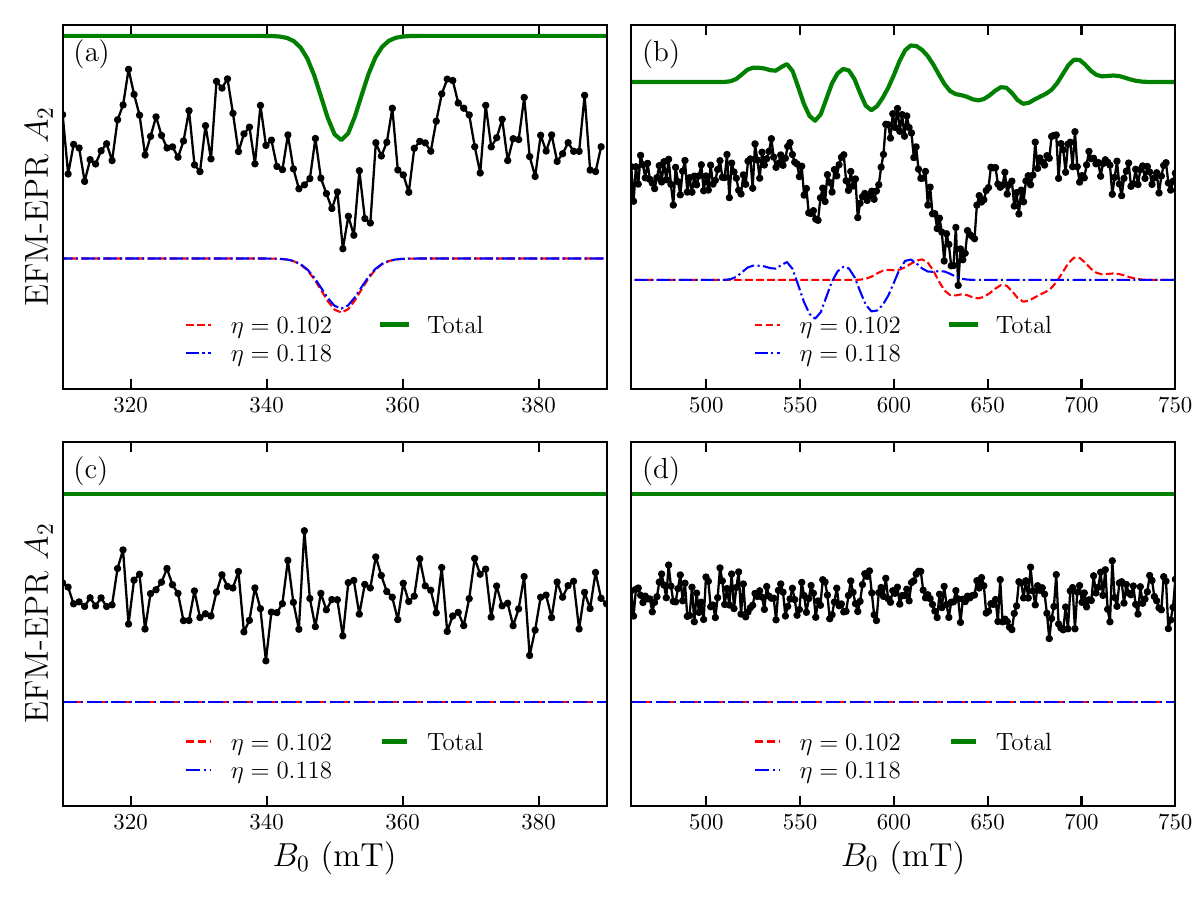}
	\caption{Comparison of the second harmonic EFM-EPR data with the multi-conformational model and (a) $\vb*{\mathcal{E}}\perp\vu{z}_M$, $\vb{B}_0\parallel\vu{z}_M$, (b) $\vb*{\mathcal{E}}\perp\vu{z}$, $\vb{B}_0\perp\vu{z}_M$, (c) $\vb*{\mathcal{E}}\parallel\vu{z}_M$, $\vb{B}_0\parallel\vu{z}_M$ and (d) $\vb*{\mathcal{E}}\parallel\vu{z}_M$, $\vb{B}_0\perp\vu{z}_M$. Simulated line shapes have been vertically offset for clarity. In all simulations, $\omega_1/2\pi = 9.74$~GHz, and $\Gamma/2\pi = 0.08$~GHz.}
	\label{fig:Experimental_Comparison}
\end{figure}

Good agreement is found between the experimental and simulated second harmonic EFM-EPR signals, for all four considered field geometries (Fig.~\ref{fig:Experimental_Comparison}). The simulations are based on the set of spin-Hamiltonian parameters that was used to model the CW-EPR first harmonic signal in Ref.~\cite{georgopoulou_dynamic_2017}: $J = 43$~cm$^{-1}$, $G_z = 4$~cm$^{-1}$, $\varphi_n=n\pi/6,$ ($n=1,\dots,12$), $\eta_1=0.102$ and $\eta_2 = 0.118$. For situations in which $\vb*{\mathcal{E}}\perp\vu{z}_M$, the simulations were conducted with a vector $\vb*{\mathcal{E}}$ that forms an angle of $17.3^\circ$ with $\vu{x}_M$, since this is the angle between $\vu{x}_M$ and the crystallographic $a$-axis [Fig.~\ref{fig:Fe3_Crystal_Simple}(a)]. No adjustment in the Hamiltonian parameters has been introduced in order to reproduce the present set of experimental results: this makes the excellent agreement between these and the simulated quantities even more significant.
Additionally, the simulation of the second-harmonic contribution predicts that the signal with $\vb{B}_0\perp\vu{z}_M$ should be $\sim 80$ times more intense than the signal with $\vb{B}_0\parallel\vu{z}_M$. This is consistent with the markedly higher signal-to-noise ratio in the data presented in Fig.~\ref{fig:Experimental_Comparison}(b), as compared to Fig.~\ref{fig:Experimental_Comparison}(a), though a detailed comparison of intensities is a complex exercise and is beyond the scope of the current work.

The lack of second-harmonic signal observed when $\vb*{\mathcal{E}}\parallel\vu{z}_M$ is further evidence that the model used here is appropriate. As mentioned previously, if the electric-field dependence of the isotropic exchange interaction is the operative spin-electric coupling, then no second-harmonic signal should be observed when $\vb*{\mathcal{E}}\parallel\vu{z}_M$, as shown in Fig.~\ref{fig:Experimental_Comparison}(c,d).

\subsection{Possible origin of the first harmonic response}
\label{subsec:Adressing_First_Harm}
The presence of a first-harmonic contribution observed in the EFM-EPR for all four geometries cannot be explained by the present model and is, more generally, incompatible with the nominal inversion symmetry of the system. There are multiple possible explanations for this first harmonic response, which we briefly discuss hereafter.

Firstly, it has been argued that the appearance of first-harmonic EFM-EPR signals in centrosymmetric systems can be attributed to spurious magnetic fields created between the electrodes by the time-dependent electric field \cite{tacconi_sensitive_2026}. In particular, according to the Ampère-Maxwell equation, time-dependent electric fields are associated with displacement currents. This mechanism can take place in any material (irrespective of its symmetry), and thus needs to be considered in order to assess the presence of a genuine EFM in EPR spectra. As detailed in the Supporting Information, the signal to noise ratio of the first harmonic signals we observe is on the order of $10^5$ times larger than the one that the spurious magnetic fields could produce. This leads us to consider other possible avenues. 

The most enticing explanation for the non-vanishing first harmonic response is some degree of correlation between the individual clusters. While the presence of inter-cluster interaction cannot \textit{per se} break inversion symmetry, such a lowering of the symmetry might be possible if the interactions result in the creation of a long-range order. Two plausible mechanisms for such correlations are the electric-dipole and the magnetic-dipole interactions.

Firstly, as was previously mentioned, there is evidence that the relatively large electric dipoles located on the pyridine molecules freeze into a disordered, glass-like state at low temperature in $\bf{Fe_3}$~\cite{Singh2025Electrical}. It is reasonable to assume that the correlation length of these dipoles should be non-zero. Additionally, the local electric field caused by these dipoles could readily interact with the spin-electric polarization of the spin triangles. Therefore, if there is semi-local ordering of pyridine dipoles, it stands to reason that the correlations in the spin triangle's polarizations should also have finite extent. To investigate this possibility, we carried out dielectric studies to determine any ferroelectric order. The $P$-$E$ plots shown as Supporting Information indicate no opening corresponding to a ferroelectric order between 300 and 20 K, which rules out the possibility of pure ferroelectric order.

Secondly, the magnetic-dipole interaction allows for an obvious inter-molecular interaction pathway. It is assumed that the strength of this interaction is too small to be important even at the 5~K temperature probed in this study. However, if the magnetic dipole interactions lead to a long-range order at zero temperature that involves a different spin configuration on $A$- and $B$-clusters, then a first harmonic EFM-EPR response could emerge. Therefore, we do not disregard the possibility that such an interaction could cause a slight biasing effect at the finite temperatures probed in this study, since only a small difference in the statistical distribution of $A$ and $B$ clusters would be needed to cause a non-vanishing first harmonic signal.

Another possible explanation for a first harmonic response is the spatial inhomogeneity of the applied fields. In fact, the shape of the crystals probed in this study ensures that the internal electric field will be non-uniform. And if the random distribution of electric dipoles is such that an $A$-cluster with a certain $\varphi$ does not have a partner $B$-cluster with $\varphi + \pi$ within a certain radius, then this pair will lead to a non-vanishing first harmonic response, since the inversion partners will not perfectly cancel. However, we expect that if this were the operative mechanism in our material, then the lineshapes of the first-harmonic responses would mimic those calculated in Section~\ref{subsec:Simulating}, and this is clearly not the case for $\vb{B}_0\parallel\vu{z}$. A similar argument holds for both thermal gradients and static magnetic-field gradients. However, we have no reason to believe that such spatial inhomogeneity would be as significant as that affecting the electric field, since magnetic shape anisotropy (demagnetizing effects) should be negligible in $\bf{Fe_3}$.

Finally, a highly probable cause of the observed first harmonic signals can be attributed to the surfaces, stacking faults or other defects that exist in the crystals. Spin triangles that exist on the surface of the crystal, or which are near stacking-faults or domain boundaries, will have greatly reduced symmetry as compared to the bulk environment. This symmetry reduction will lead to additional terms in the spin Hamiltonian, including in-plane DMI interactions for example. These, when paired with the mechanism described in Ref.~\cite{Katsura2005-qs}, could give rise to a ``derivative-like'' EFM-EPR response for $\vb*{\mathcal{E}}\parallel\vu{z}$. This mechanism would lead to a first harmonic signal that is proportional to the number of spin-triangles situated in these low-symmetry environments, which should be small relative to the bulk. This would be consistent with the observation that the second-harmonic signal, dominated by the bulk contribution, can be successfully simulated through a spin Hamiltonian that does not include these low-symmetry terms. A more detailed discussion of this effect is included in the Supporting Information.


An applied electric field can in principle affect not only the isotropic exchange in this system, but also the DMI, the $g$-tensors, and the single-ion anisotropies. Symmetry arguments and previous experiments on related systems have shown that an applied longitudinal electric field ($\vb{E}_\mathrm{ext}\parallel\vb{z}$) can renormalize the in-plane components of the DM couplings $\vb{G}_{ij}$, just like a transverse electric field renormalizes the isotropic exchange couplings $J_{ij}$. However, 
the fact that the second-harmonic signal observed in both longitudinal and transverse static magnetic fields agrees so well with the response predicted by the current model, which has already been validated by previous experiments \cite{leMardele25a}, justifies the assumption that the modulation of isotropic exchange is the dominant spin-electric coupling mechanism in the $\bf{Fe_3}$ molecules.

\section{Conclusions}
\label{sec:Conclusions}
The primary conclusion to be drawn from this work is that $\bf{Fe}_3$ single crystals exhibit a clear second harmonic EFM-EPR response, which is consistent with the expectation that the primary spin-electric coupling mechanism is a modulation of the isotropic exchange interaction. The observed second-harmonic signal was simulated through a previously validated model, without ad hoc modifications of the physical parameters. The model accounts for a symmetry-breaking mechanism, related to Jahn-Teller distortions of the spin triangle, which allows, at the single-molecule level, a linear electric-field induced modulation of the energy levels.

In addition to the expected second harmonic response, a clear first harmonic response was observed in all the considered experimental setups. We hypothesize that the first harmonic signal is largely due to a small subensemble of molecules, located in the vicinity of surfaces or defects (stacking faults, vacancies, etc.). In such environments, the local symmetry can be drastically reduced as compared to the bulk, leading to otherwise forbidden terms in both the zero-field and electric-field-dependent spin-Hamiltonian. The first-harmonic signal might also be due inter-cluster interactions, resulting in the formation of longer-range magnetic correlations and spontaneous symmetry breaking. A detailed analysis of both the magnetic- and the electric-dipole interactions in the system will be needed in order to further explore this possibility.

Methodologically, the present study suggests that EFM-EPR measurements present a wider than expected range of applicability including its first-harmonic variant applied on centrosymmetric crystals. In particular, whereas the first-harmonic signal should be unrelated with sites sharing the crystal's nominal centrosymmetry, it should account for sites near crystal faces and imperfections. It would therefore reveal underlying ME phenomena prior to their being quenched by a centrosymmetric crystal arrangement. In the domain of molecular magnetism, where the question of interest lies with the intrinsic properties of isolated molecules (i.e. in the context of nanodevices), such a signature would also be+65 of great interest.

\section*{Acknowledgements}
The authors acknowledge financial support from the French Agence Nationale de la Recherche (ANR), under Grant No. ANR-24-CE29-2752-01 (project SPINCHIRAL). The authors thank Simon Schmitt (Faculty of Physics and Engineering of the University of Strasbourg) for help with the construction of the sample holder.

\section*{Supporting information}
The following files are available free of charge.
\begin{itemize}
  \item Supporting-Information: Detailed $\varphi$-dependence of thermal expectation values $\left<\vb*{\mu}\right>$, $\left<\vb{p}\right>$ and $\left<E\right>$, a calculation of the expected time-varying electric field-induced spurious magnetic field strengths in our experimental setup, a description of the dielectric properties of $\bf{Fe_3}$ and a brief discussion of symmetry reductions near defects in the crystal.
\end{itemize}

\printbibliography

@article{Boudalis18a,
author = {Boudalis, Athanassios K. and Robert, Jérôme and Turek, Philippe},
title = {First Demonstration of Magnetoelectric Coupling in a Polynuclear Molecular Nanomagnet: Single-Crystal {EPR} Studies of [{F}e3{O}({O}2{CP}h)6(py)3]{C}l{O}4$\cdot$py under Static Electric Fields},
journal = {Chemistry – A European Journal},
volume = {24},
number = {56},
pages = {14896-14900},
doi = {https://doi.org/10.1002/chem.201803038},
url = {https://chemistry-europe.onlinelibrary.wiley.com/doi/abs/10.1002/chem.201803038},
year = {2018}
}

@article{Bulaevskii08a,
  title = {Electronic orbital currents and polarization in Mott insulators},
  author = {Bulaevskii, L. N. and Batista, C. D. and Mostovoy, M. V. and Khomskii, D. I.},
  journal = {Phys. Rev. B},
  volume = {78},
  issue = {2},
  pages = {024402},
  numpages = {9},
  year = {2008},
  month = {Jul},
  publisher = {American Physical Society},
  doi = {10.1103/PhysRevB.78.024402},
  url = {https://link.aps.org/doi/10.1103/PhysRevB.78.024402}
}

@Article{leMardele25a,
author={le Mardel{\'e}, Florian
and Mohelsk{\'y}, Ivan
and Wyzula, Jan
and Orlita, Milan
and Turek, Philippe
and Troiani, Filippo
and Boudalis, Athanassios K.},
title={Probing spin-electric transitions in a molecular exchange qubit},
journal={Nature Communications},
year={2025},
month={Jan},
day={30},
volume={16},
number={1},
pages={1198},
issn={2041-1723},
doi={10.1038/s41467-025-56453-1},
url={https://doi.org/10.1038/s41467-025-56453-1}
}

@article{Perfetti26a,
  author = "Perfetti, Mauro and Sessoli, Roberta",
  title = "Electric-Field Control of Molecular Spins for Quantum Information", 
  journal= "Annual Review of Physical Chemistry",
  year = "2026",
  volume = "77",
  number = "Volume 77, 2026",
  pages = "443-464",
  doi = "https://doi.org/10.1146/annurev-physchem-082624-102024",
  url = "https://www.annualreviews.org/content/journals/10.1146/annurev-physchem-082624-102024",
  publisher = "Annual Reviews",
  issn = "1545-1593",
  type = "Journal Article"
 }

@article{Trif08a,
  title = {Spin-Electric Coupling in Molecular Magnets},
  author = {Trif, Mircea and Troiani, Filippo and Stepanenko, Dimitrije and Loss, Daniel},
  journal = {Phys. Rev. Lett.},
  volume = {101},
  issue = {21},
  pages = {217201},
  numpages = {4},
  year = {2008},
  month = {Nov},
  publisher = {American Physical Society},
  doi = {10.1103/PhysRevLett.101.217201},
  url = {https://link.aps.org/doi/10.1103/PhysRevLett.101.217201}
}

@article{Trif10a,
  title = {Spin electric effects in molecular antiferromagnets},
  author = {Trif, Mircea and Troiani, Filippo and Stepanenko, Dimitrije and Loss, Daniel},
  journal = {Phys. Rev. B},
  volume = {82},
  issue = {4},
  pages = {045429},
  numpages = {29},
  year = {2010},
  month = {Jul},
  publisher = {American Physical Society},
  doi = {10.1103/PhysRevB.82.045429},
  url = {https://link.aps.org/doi/10.1103/PhysRevB.82.045429}
}

@article{Troiani12a,
  title = {Hyperfine-induced decoherence in triangular spin-cluster qubits},
  author = {Troiani, Filippo and Stepanenko, Dimitrije and Loss, Daniel},
  journal = {Phys. Rev. B},
  volume = {86},
  issue = {16},
  pages = {161409},
  numpages = {5},
  year = {2012},
  month = {Oct},
  publisher = {American Physical Society},
  doi = {10.1103/PhysRevB.86.161409},
  url = {https://link.aps.org/doi/10.1103/PhysRevB.86.161409}
}

@book{Mims1976linear,
title = {The linear electric field effect in paramagnetic resonance},
author = {Mims, W. B.},
address = {Oxford},
year = {1976},
publisher = {Clarendon Press}
}

@article{Singh2025Electrical,
	author = {Singh Chauhan, Balwant and Chatterjee, Ratnamala and Turek, Philippe and Boudalis, Athanassios K.},
	journal = {Journal of the American Chemical Society},
	doi = {10.1021/jacs.5c05601},
	issn = {0002-7863, 1520-5126},
	number = {23},
	year = {2025},
	month = {jun 11},
	note = {publisher: American Chemical Society (ACS)},
	pages = {20063--20070},
	title = {Electrical {Detection} of {Magnetic} {Perturbations} through the {Magnetoelectric} {Effect} of a {Molecular} {Spin} {Triangle}},
	volume = {147},
}

@article{maisuradze_magnetoelectric_2012,
	title = {Magnetoelectric {Coupling} in {Single} {Crystal} {Cu}$_{\textrm{2}}${OSeO}$_{\textrm{3}}$ {Studied} by a {Novel} {Electron} {Spin} {Resonance} {Technique}},
	volume = {108},
	copyright = {http://link.aps.org/licenses/aps-default-license},
	issn = {0031-9007, 1079-7114},
	url = {https://link.aps.org/doi/10.1103/PhysRevLett.108.247211},
	doi = {10.1103/PhysRevLett.108.247211},
	language = {en},
	number = {24},
	urldate = {2025-02-05},
	journal = {Physical Review Letters},
	author = {Maisuradze, A. and Shengelaya, A. and Berger, H. and Djokić, D. M. and Keller, H.},
	month = jun,
	year = {2012},
	pages = {247211},
}

@article{laguta_magnetoelectric_2020,
	title = {Magnetoelectric coupling in multiferroic {Z}-type hexaferrite revealed by electric-field-modulated magnetic resonance studies},
	volume = {55},
	issn = {0022-2461, 1573-4803},
	url = {http://link.springer.com/10.1007/s10853-020-04563-0},
	doi = {10.1007/s10853-020-04563-0},
	language = {en},
	number = {18},
	urldate = {2026-05-08},
	journal = {Journal of Materials Science},
	author = {Laguta, Valentin and Kempa, Martin and Bovtun, Viktor and Buršík, Josef and Zhai, Kun and Sun, Young and Kamba, Stanislav},
	month = jun,
	year = {2020},
	pages = {7624--7633},
}

@article{kita_electric_1979,
	title = {Electric {Shift} in the {Antiferromagnetic} {Resonance} and the {Mechanism} of the {Parallel} {Magnetoelectric} {Effect} of {Cr}$_{\textrm{2}}${O}$_{\textrm{3}}$},
	volume = {46},
	issn = {0031-9015, 1347-4073},
	url = {https://journals.jps.jp/doi/10.1143/JPSJ.46.1033},
	doi = {10.1143/JPSJ.46.1033},
	language = {en},
	number = {3},
	urldate = {2026-05-08},
	journal = {Journal of the Physical Society of Japan},
	author = {Kita, Eiji and Siratori, Kiiti and Tasaki, Akira},
	month = mar,
	year = {1979},
	pages = {1033--1034},
}

@article{tokura_multiferroics_2014,
	title = {Multiferroics of spin origin},
	volume = {77},
	issn = {0034-4885, 1361-6633},
	url = {http://stacks.iop.org/0034-4885/77/i=7/a=076501?key=crossref.e08f6ce66f85d4c81a3467011ab4b52a},
	doi = {10.1088/0034-4885/77/7/076501},
	number = {7},
	urldate = {2019-06-19},
	journal = {Reports on Progress in Physics},
	author = {Tokura, Yoshinori and Seki, Shinichiro and Nagaosa, Naoto},
	month = jul,
	year = {2014},
	pages = {076501},
}

@article{song_exploring_2025,
	title = {Exploring {Spin}‐{Electric} {Coupling} in an {Electrically}‐{Controlled} {Rare}‐{Earth} {Molecular} {Qubit}},
	volume = {64},
	issn = {1433-7851, 1521-3773},
	url = {https://onlinelibrary.wiley.com/doi/10.1002/anie.202513081},
	doi = {10.1002/anie.202513081},
	language = {en},
	number = {40},
	urldate = {2026-05-08},
	journal = {Angewandte Chemie International Edition},
	author = {Song, Ji‐Min and Chen, Jia‐Xin and Zhang, Yu‐Shuang and Deng, Qing‐Song and Xie, Yi and Gao, Song and Wang, Ye‐Xin and Liu, Zheng and Jiang, Shang‐Da},
	month = sep,
	year = {2025},
	pages = {e202513081},
}

@article{tacconi_sensitive_2026,
	title = {Sensitive detection of spin-electric coupling in a {Cr}$_{\textrm{3}}$ antiferromagnetic triangle},
	volume = {17},
	issn = {2041-6520, 2041-6539},
	url = {https://xlink.rsc.org/?DOI=D5SC08012F},
	doi = {10.1039/D5SC08012F},
	language = {en},
	number = {6},
	urldate = {2026-05-08},
	journal = {Chemical Science},
	author = {Tacconi, Leonardo and Bisht, Shubham and Cini, Alberto and Perfetti, Mauro and Orlando, Tomas and Fittipaldi, Maria and Shatruk, Michael and Sessoli, Roberta},
	year = {2026},
	pages = {3329--3338},
}

@article{georgopoulou_dynamic_2017,
	title = {Dynamic versus {Static} {Character} of the {Magnetic} {Jahn}–{Teller} {Effect}: {Magnetostructural} {Studies} of [{Fe}$_{\textrm{3}}${O}({O}$_{\textrm{2}}${CPh})$_{\textrm{6}}$(py)$_{\textrm{3}}$]{ClO}$_{\textrm{4}}$·py},
	volume = {56},
	issn = {0020-1669, 1520-510X},
	shorttitle = {Dynamic versus {Static} {Character} of the {Magnetic} {Jahn}–{Teller} {Effect}},
	url = {http://pubs.acs.org/doi/abs/10.1021/acs.inorgchem.6b01912},
	doi = {10.1021/acs.inorgchem.6b01912},
	language = {en},
	number = {2},
	urldate = {2017-01-06},
	journal = {Inorganic Chemistry},
	author = {Georgopoulou, Anastasia N. and Margiolaki, Irene and Psycharis, Vassilis and Boudalis, Athanassios K.},
	month = jan,
	year = {2017},
	pages = {762--772},
}

@article{murao_jahn-teller_1974,
	title = {Jahn-{Teller} effect in trinuclear complexes},
	volume = {49},
	issn = {03759601},
	url = {http://linkinghub.elsevier.com/retrieve/pii/0375960174906574},
	doi = {10.1016/0375-9601(74)90657-4},
	language = {en},
	number = {1},
	urldate = {2016-05-26},
	journal = {Physics Letters A},
	author = {Murao, T.},
	month = aug,
	year = {1974},
	pages = {33--35},
}

@ARTICLE{Moriya1960-xc,
  title     = "Anisotropic Superexchange Interaction and Weak Ferromagnetism",
  author    = "Moriya, T{\^o}ru",
  journal   = "Physical Review",
  publisher = "American Physical Society (APS)",
  volume    =  120,
  number    =  1,
  pages     = "91--98",
  month     =  oct,
  year      =  1960,
  copyright = "http://link.aps.org/licenses/aps-default-license"
}

@ARTICLE{Sowrey2001-oz,
  title     = "Spin frustration and concealed asymmetry: structure and magnetic
               spectrum of [{Fe3O(O2CPh)6(py)3]ClO4·py}",
  author    = "Sowrey, Frank E and Tilford, Claire and Wocadlo, Sigrid and
               Anson, Christopher E and Powell, Annie K and Bennington, Stephen
               M and Montfrooij, Wouter and Jayasooriya, Upali A and Cannon,
               Roderick D",
  journal   = "J. Chem. Soc.",
  publisher = "Royal Society of Chemistry (RSC)",
  number    =  6,
  pages     = "862--866",
  year      =  2001
}

@ARTICLE{Rakitin1981-bv,
  title     = "{EPR} spectra of trigonal clusters",
  author    = "Rakitin, Yu V and Yablokov, Yu V and Zelentsov, V V",
  journal   = "J. Magn. Reson.",
  publisher = "Elsevier BV",
  volume    =  43,
  number    =  2,
  pages     = "288--301",
  month     =  may,
  year      =  1981,
  language  = "en"
}

@ARTICLE{Ham1961-te,
  title     = "Linear effect of applied electric field in electron spin
               resonance",
  author    = "Ham, Frank S",
  journal   = "Phys. Rev. Lett.",
  publisher = "American Physical Society (APS)",
  volume    =  7,
  number    =  6,
  pages     = "242--243",
  month     =  sep,
  year      =  1961,
  copyright = "http://link.aps.org/licenses/aps-default-license",
  language  = "en"
}

@ARTICLE{Ludwig1961-sr,
  title     = "Splitting of electron spin resonance lines by an applied
               electric field",
  author    = "Ludwig, G W and Woodbury, H H",
  journal   = "Phys. Rev. Lett.",
  publisher = "American Physical Society (APS)",
  volume    =  7,
  number    =  6,
  pages     = "240--241",
  month     =  sep,
  year      =  1961,
  copyright = "http://link.aps.org/licenses/aps-default-license",
  language  = "en"
}

@ARTICLE{Katsura2005-qs,
  title     = "Spin current and magnetoelectric effect in noncollinear magnets",
  author    = "Katsura, Hosho and Nagaosa, Naoto and Balatsky, Alexander V",
  journal   = "Phys. Rev. Lett.",
  publisher = "American Physical Society (APS)",
  volume    =  95,
  number    =  5,
  pages     = "057205",
  month     =  jul,
  year      =  2005,
  copyright = "http://link.aps.org/licenses/aps-default-license",
  language  = "en"
}

@ARTICLE{Azimi_Mousolou2016-lt,
  title     = "Spin-electric Berry phase shift in triangular molecular magnets",
  author    = "Azimi Mousolou, Vahid and Canali, C M and Sj{\"o}qvist, Erik",
  journal   = "Phys. Rev. B.",
  publisher = "American Physical Society (APS)",
  volume    =  94,
  number    =  23,
  month     =  dec,
  year      =  2016,
  copyright = "http://link.aps.org/licenses/aps-default-license"
}

@BOOK{Boas2005-ky,
  title     = "Mathematical methods in the physical sciences",
  author    = "Boas, M L",
  publisher = "John Wiley \& Sons",
  edition   =  3,
  month     =  jul,
  year      =  2005,
  address   = "Nashville, TN",
  language  = "en"
}


\end{document}